\documentclass[runningheads]{llncs}
\usepackage{float}
\usepackage[T1]{fontenc}
\usepackage{graphicx}
\usepackage{booktabs}
\usepackage{amsmath}
\usepackage{amssymb}
\usepackage[table]{xcolor}
\usepackage{enumitem}
\usepackage{multirow}
\usepackage{tabularx}
\usepackage{hyperref}
\usepackage{comment}
\usepackage{array}
\usepackage{placeins}
\usepackage{ragged2e}

\newcolumntype{Y}{>{\RaggedRight\arraybackslash}X}

\definecolor{HeaderBlue}{HTML}{D9EAF7}
\definecolor{PhaseGray}{HTML}{F2F2F2}

\title{
{\small\textbf{Author copy of the camera-ready paper accepted for publication in the Industry Track of the 20th European Conference on Software Architecture (ECSA 2026).}}\\[1em]
TimeArch: A Multi-Agent Approach for Software Architecture Design
}

\author{
Muhammad Waseem\inst{1} \and
Mateen Abbasi\inst{2} \and
Viljami Järvinen\inst{2} \and
Petrus Lipsanen\inst{2} \and
Navneet Arora\inst{1} \and
Kalle Kulonen\inst{1} \and
Kalle Mäkelä\inst{3} \and
Tommi Mikkonen\inst{2} \and
Pekka Abrahamsson\inst{1}
}

\authorrunning{M. Waseem et al.}

\institute{
Faculty of Information Technology and Communication Sciences, Tampere University, Tampere, Finland
\email{\{muhammad.waseem, navneet.arora, kalle.kulonen, pekka.abrahamsson\}@tuni.fi}
\and
University of Jyväskylä, Jyväskylä, Finland
\email{\{mateen.a.abbasi, viljami.a.e.jarvinen, petrus.i.lipsanen, tommi.j.mikkonen\}@jyu.fi}
\and
Aimbition Oy, Finland
\email{kalle.makela@aimbition.com}
}

\begin{document}

\maketitle

\begin{abstract}
Generative AI is making software implementation faster, but not necessarily better architected. Most current GenAI coding workflows move directly from prompts to code, while the architectural reasoning that should connect requirements, quality attributes, stakeholders, design decisions, and implementation remains weakly supported. Architectural styles, system decomposition, tradeoff analysis, quality-attribute decisions, and stakeholder concerns are often left implicit or handled outside the AI-supported workflow. This creates a missing architectural layer between requirements and code, limiting traceability, validation, and governance in GenAI-enabled software development.

This paper presents \textbf{TimeArch}, a multi-agent approach that makes this architectural layer explicit. TimeArch structures software architecture design as a governed lifecycle rather than a one-shot generation task. It combines agent-based reasoning with retrieval-augmented grounding to support requirement definition, architecture design, validation and assurance, and delivery and evolution. The approach supports requirement capture, architectural driver identification, design reasoning, validation, and decision documentation within a traceable process.

We demonstrate TimeArch through a working prototype that operationalizes the proposed approach and assess its feasibility using representative software architecture design cases from multiple domains with heterogeneous inputs. 
As future work, we plan to conduct comprehensive industrial case studies to evaluate the adoption and use of TimeArch in both new and existing software projects.

A TimeArch demo is available at: \url{https://sda-assistant.com/}.


\end{abstract}
\section{Introduction}

Software architecture design is a decision-intensive engineering activity that must align business goals, stakeholder concerns, system constraints, and quality attributes within a consistent architectural structure~\cite{bass2012}. 
It often relies on heterogeneous artifacts, manually maintained architectural decisions, and loosely connected tools, which limits traceability between requirements, architectural decisions, and implementation~\cite{capilla2016}. 
Over time, this can lead to divergence between the intended architecture and the implemented system, a phenomenon widely recognized as architectural erosion~\cite{desilva2012}.

\textbf{Context and Challenges}: Recent advances in Large Language Models (LLMs) have improved the automation of code generation and developer assistance tasks~\cite{chen2021codex,Ebert2023GenerativeAI}. 
These models are increasingly used to generate implementation artifacts from informal or partially structured requirements, often with limited explicit consideration of architectural reasoning, design alternatives, and decision rationale~\cite{waseem2025vibe}. 
This trend is reflected in emerging AI-assisted development platforms, where requirements can be translated into code with limited support for lifecycle-level architecture design and governance~\cite{ernst2022ai}. 
Consequently, such approaches operate primarily at the implementation level and provide limited mechanisms for capturing architectural intent, evaluating design alternatives, documenting architectural rationale, and maintaining consistency across related artifacts~\cite{Ozkaya2023ArchitectureKnowledgeGenAI}. 
As a result, architecture may become implicitly embedded in generated code rather than explicitly modeled, validated, or governed.

\textbf{Research Gap}: Existing architecture design methods provide disciplined processes for reasoning about requirements, quality attributes, design alternatives, trade-offs, and architecture decisions. 
However, these methods are not yet well integrated into GenAI-enabled software development workflows. 
Current GenAI coding workflows rarely provide a governed architecture lifecycle, while existing AI-assisted and multi-agent approaches tend to focus on isolated software engineering (SE) tasks. 
They provide limited support for traceability, validation, and governance across the transition from requirements to architecture and implementation. 
This leaves an architectural gap between requirements-level intent and implementation-level generation in GenAI-enabled development.

\textbf{Motivation}: These limitations highlight a broader challenge in AI-assisted SE: generative AI can accelerate the production of software artifacts, but architecture-related concerns remain insufficiently represented in the process. 
Without explicit architectural reasoning, decision documentation, and governance mechanisms, the transition from requirements to implementation risks becoming difficult to trace, validate, and control. 
This motivates approaches that integrate generative AI with structured architecture design methods. 
Accordingly, this work aims to establish a lifecycle-oriented and human-governed approach that supports traceability, validation, and controlled progression from requirements to implementation.

To address this gap, we propose \textbf{TimeArch}, a multi-agent approach for requirements-driven software architecture design. TimeArch structures architecture design around four lifecycle phases: requirement definition, architecture design, validation and assurance, and delivery and evolution. These phases are supported by LLM-based generator and evaluator agents with explicit governance checkpoints to produce, assess, refine, and approve architectural artifacts.



The primary contributions of this paper are:
\begin{itemize}
\item A lifecycle-oriented multi-agent approach for requirements-driven software architecture design that structures architecting across requirement definition, architecture design, validation and assurance, and delivery and evolution.
\item A governance and traceability mechanism that combines agent-generated artifacts, evaluator checks, human approval checkpoints, and audit-oriented documentation to support controlled architectural reasoning.
\item A prototype implementation and case-based feasibility evaluation demonstrating how the approach can structure requirements, derive architectural drivers, generate design artifacts, produce validation reports, and consolidate architecture documentation across representative software systems.
\end{itemize}


\section{Proposed Approach}
\label{ProposedApproach}

We present \textbf{TimeArch}, a multi-agent approach for software architecture design. 
TimeArch structures architectural work as a lifecycle rather than as a single design step, reflecting established views of architecture as an iterative process involving requirements, design decisions, evaluation, documentation, and evolution~\cite{bass2012,cervantes2016designing,clements2002evaluating}. 
The approach is grounded in the SEI Attribute-Driven Design (ADD~3.0) method~\cite{cervantes2016designing}, the ISO/IEC~25010 quality model~\cite{iso25010}, and the AWS Well-Architected Framework~\cite{awsWellArchitected}. 
These foundations guide agent reasoning, artifact evaluation, and decision documentation, and provide the basis for maintaining traceability and auditability in architecture knowledge management~\cite{capilla2016}.

The main contribution of TimeArch is to operationalize established architecture design and evaluation knowledge through coordinated generator and evaluator agents under human governance. 
Instead of relying on unconstrained prompt-based generation, TimeArch organizes architectural work around phase-specific agent activities, explicit evaluation criteria, a shared knowledge base, and approval gates. 
The human architect remains responsible for reviewing outputs, resolving tradeoffs, accepting risks, and approving progression across lifecycle stages.

\begin{figure}
    \centering
    \includegraphics[width=1\linewidth]{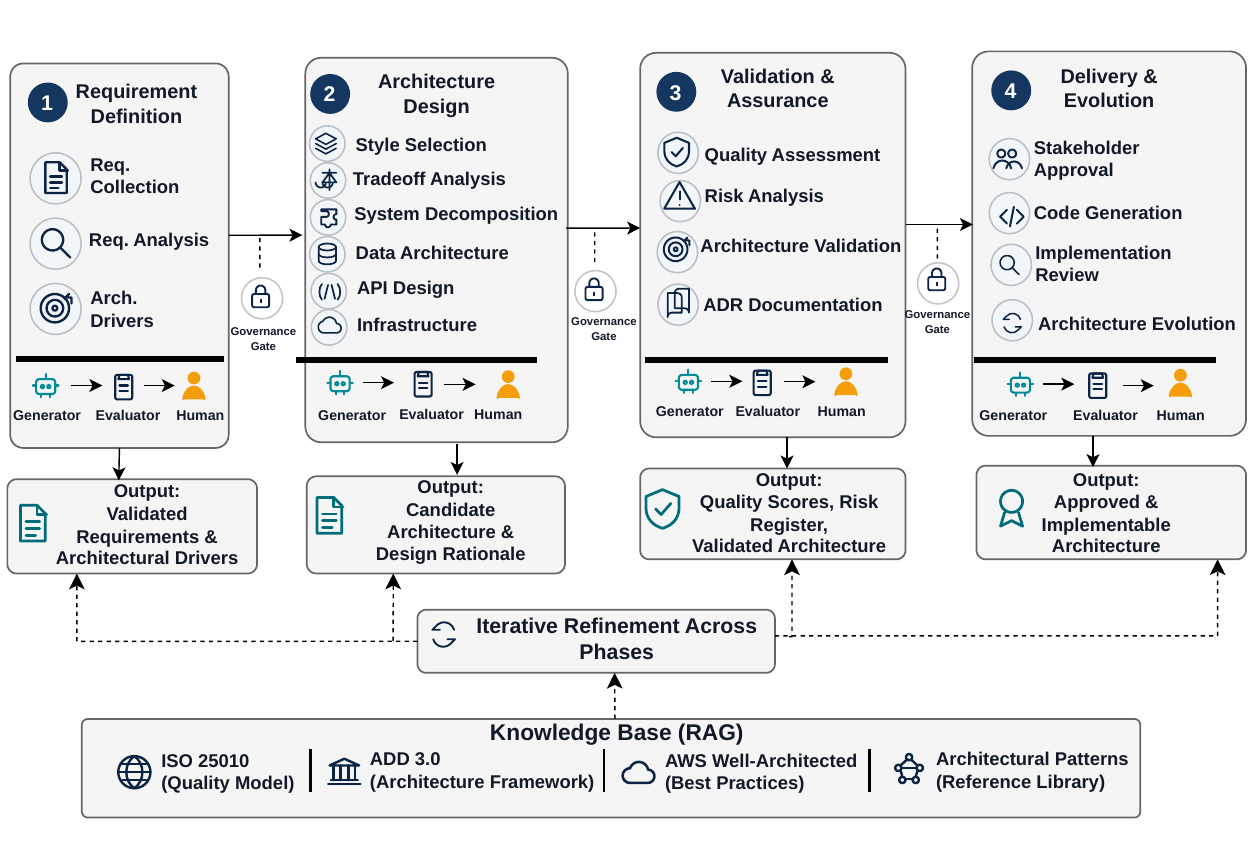}
    \caption{Overview of the proposed approach.}
    \label{fig:ApproachOverview}
\end{figure}

Figure~\ref{fig:ApproachOverview} provides an overview of TimeArch. 
The lifecycle consists of four phases: Requirement Definition, Architecture Design, Validation and Assurance, and Delivery and Evolution. 
Across these phases, generated artifacts, evaluation results, architectural decisions, risks, feedback, and traceability links are maintained in a shared knowledge base to support continuity and refinement.

\subsection{Agent Roles and Governance}

TimeArch follows a \emph{generator--evaluator--human} control pattern.
Generator agents create architecture-related artifacts, evaluator agents assess them against explicit criteria, and the human architect makes the final decision.
Across the lifecycle, agents support requirement extraction, architectural driver identification, candidate architecture generation, tradeoff analysis, ADR preparation, validation, implementation guidance, and evolution recording.
A detailed description of the agent roles and responsibilities is available on the TimeArch prototype landing page.

Governance is implemented through approval gates between phases.
A phase can progress only when mandatory artifacts are available, critical evaluator findings are resolved or accepted, traceability links are complete, and the human architect approves the output.
If unresolved issues remain, the artifact is returned for revision and re-evaluation.
This gate-based control keeps architectural risks, rationale, assumptions, and decisions explicit while ensuring that TimeArch assists architectural work without transferring decision authority from the human architect to agents.


\subsection{TimeArch Lifecycle}
TimeArch is organized into four lifecycle stages, described below.

\textbf{Phase 1 -- Requirement Definition}:
This phase establishes the input baseline for subsequent architectural reasoning~\cite{iso29148,bass2012}.
Generator agents extract functional requirements, non-functional requirements, constraints, assumptions, integrations, actors, business rules, and stakeholder concerns from project inputs~\cite{iso29148,incose2017}.
Non-functional requirements are normalized using ISO/IEC~25010~\cite{iso25010}, while requirement quality is checked using ISO/IEC/IEEE~29148 and INCOSE-style criteria~\cite{iso29148,incose2017}.
The phase produces validated requirements, architectural drivers, quality attribute mappings, assumptions, constraints, and traceability links.

\textbf{Phase 2 -- Architecture Design}:
This phase transforms validated requirements and architectural drivers into a candidate architecture specification~\cite{bass2012,cervantes2016designing}.
Following ADD~3.0-style reasoning, generator agents support style selection, decomposition, tradeoff analysis, data ownership, interface and API design, cross-cutting concern analysis, and infrastructure planning~\cite{cervantes2016designing,evans2003ddd}.
Evaluator agents assess the candidate architecture against quality attributes, constraints, design consistency, feasibility, and traceability~\cite{iso25010,iso42010}.
The phase produces component models, tradeoff reports, API contracts, data ownership maps, infrastructure options, and design rationale~\cite{iso42010,kruchten1995architectural}.

\textbf{Phase 3 -- Validation and Assurance}:
This phase evaluates the candidate architecture before delivery using established architecture evaluation practices~\cite{kazman2000atam,clements2002evaluating}.
It checks quality-attribute satisfaction, risks and mitigations, traceability, rationale consistency, and ADR completeness~\cite{iso25010,capilla2016}.
The validation logic is grounded in ATAM, architecture quality models, and architecture knowledge management practices.
The phase produces a validated architecture package, quality assessment results, a risk register, mitigation notes, ADRs, and required correction items.

\textbf{Phase 4 -- Delivery and Evolution}:
This phase governs the transition from approved architecture to implementation and supports the preservation of architectural intent over time~\cite{bass2012,clements2002evaluating}.
TimeArch does not replace implementation processes; rather, it provides architecture-aligned guidance and conformance checks for generated or manually implemented artifacts~\cite{Ozkaya2023ArchitectureKnowledgeGenAI,Ebert2023GenerativeAI}.
In this phase, implementation artifacts and proposed changes are examined for alignment with the approved architecture baseline and for potential architectural drift, erosion, or technical debt~\cite{desilva2012,waseem2025vibe}.
The phase produces an approved architecture baseline, implementation guidance, conformance evidence, change records, and evolution monitoring results.

\section{System Architecture and Implementation}
\label{SystemArchitectureandImplementation}

To realize the proposed approach, we developed \textbf{TimeArch} as a live prototype. 


\begin{figure}[H]
    \centering
    \includegraphics[width=1\linewidth]{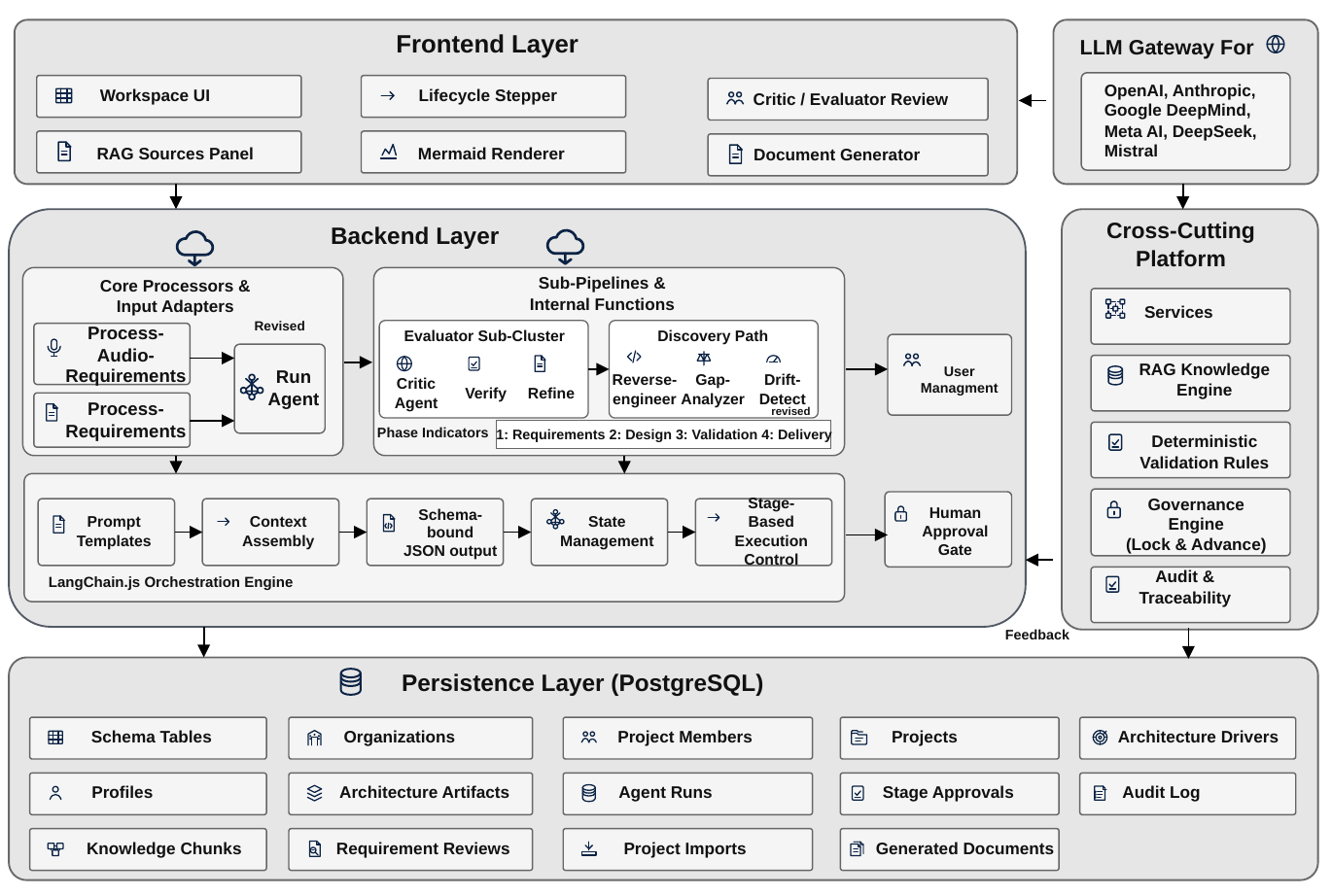}
    \caption{High-level architecture of the TimeArch prototype.}
    \label{fig:Architecture}
\end{figure}
\subsection{System Architecture Overview}
 
Figure~\ref{fig:Architecture} shows the high-level architecture of the TimeArch prototype. The system follows a layered design that separates user interaction, backend services, orchestration, multi-agent execution, cross-cutting services, and persistence. 
The \emph{frontend layer} provides stage-oriented workspaces for requirement intake, lifecycle execution, artifact inspection, diagram rendering, and document generation. 
These workspaces map to the four phases of the proposed approach. 
The \emph{backend layer} is implemented as serverless edge functions that handle execution requests, context preparation, persistence, and access control.

At the center of the system is an \emph{orchestration engine} built on LangChain.js. 
It coordinates execution across lifecycle stages, assembles stage-specific context, manages state, binds tools for structured output, and enforces stage-based execution control. 
The \emph{multi-agent lifecycle pipeline} groups specialized agents into the four phases of the approach: requirement-analysis agents, design agents, validation agents, and delivery agents. 
A set of \emph{cross-cutting services}, including a RAG knowledge engine, evaluator agents, deterministic validation rules, a governance engine, and an audit and traceability service, can be invoked from different stages of the pipeline. 
Artifacts, including requirements, architectural drivers, architecture artifacts, agent runs, stage approvals, audit logs, and knowledge chunks, are stored in a PostgreSQL \emph{persistence layer}.

\subsection{Multi-Agent Orchestration and Execution}

TimeArch implements the lifecycle through specialized generator agents for requirement analysis, architecture design, validation, documentation, and delivery support. A small number of stages act as manual governance checkpoints, where critical decisions are reviewed and approved by the human architect without automated generation. For each stage, the orchestrator loads a stage-specific prompt, assembles the execution context from prior-stage artifacts, retrieves relevant knowledge chunks through the RAG engine, and constrains the LLM output using a stage-specific schema. Schema-bound outputs are validated before being persisted as architecture artifacts, which helps keep inter-stage interactions consistent and machine-readable. The lifecycle is executed as a staged pipeline. After requirement capture and validation, the orchestrator constructs an execution context that includes the validated requirements, previously generated artifacts, and relevant knowledge chunks. Each stage's output is validated against its schema, optionally evaluated by an evaluator agent, and then made available for review and progression. TimeArch uses bounded retries rather than open-ended regeneration. Transient LLM call failures are retried automatically with a fixed upper bound, and if an agent response cannot be parsed into the required structured schema, the agent is invoked at most once more with reinforced structured-output instructions. If this second attempt also fails, the run is marked as failed. Further reruns are controlled by the human-governed stage workflow: the user may revise inputs or rerun a stage before locking it, and these iterations are recorded in the agent run and audit logs.

\subsection{Prompting Strategy}
The prompting strategy decomposes architecture reasoning into narrower agent tasks instead of relying on a single general prompt. Different prompts are used for requirement analysis, architecture design, quality and validation checks, delivery support, challenger review, and refinement. This makes the agent behavior more transparent and allows prompt use to be inspected, reviewed, and updated as the prototype evolves. The prototype provides a Prompt Library where authenticated users can inspect the current set of 28 prompts in read-only mode.

\subsection{Evaluation, Refinement, and Governance}

The \emph{generator--evaluator--human} pattern is realized through evaluator agents that operate alongside generator agents. 
The requirement critic agent reviews agent-extracted requirements and architectural drivers against ISO/IEEE~29148, INCOSE writing rules, and ATAM-style soundness criteria. 
It returns structured verdicts, severity levels, rationales, violated rule codes, and suggested rewrites when needed. 
For architecture design and validation stages, a checklist verification agent reviews stage artifacts against architecturally significant concerns. 
When gaps are identified, a refinement agent can produce a focused patch that addresses the failing checklist item while preserving the rest of the artifact. 
In addition, the design phase includes a tradeoff and overengineering evaluator that critiques candidate architectures, supports Architecture Decision Record (ADR) generation, and flags overengineering risks. 
For brownfield evolution, additional agents support reverse engineering, gap analysis, and drift detection.

Validation and governance are embedded in the system to control execution and improve consistency across artifacts. 
\emph{Deterministic validation rules} verify completeness, consistency, and traceability between related artifacts. 
For example, they check whether every architectural driver traces back to at least one requirement and whether each ADR references at least one driver. 
Governance is enforced by a dedicated \emph{governance engine} that manages a per-stage lock-and-advance protocol. 
A stage can only be advanced when deterministic validations pass, evaluator verdicts are addressed, and the human architect explicitly approves it. 
Approved stages are locked, and any unlocking action is recorded.

\subsection{Knowledge Grounding}
To ground design decisions in established practice, TimeArch includes a RAG knowledge engine.
A knowledge table stores curated excerpts from architectural methods, standards, and frameworks.
These excerpts are tagged by framework, category, and lifecycle stage. A PostgreSQL full-text search function returns the top-$k$ relevant chunks for a given query and lifecycle stage, and the orchestrator injects these chunks into the agent's context window.
The retrieved chunks are also exposed to the user through a RAG sources panel, allowing generated artefacts to be inspected alongside the guidance that informed them. The current prototype uses PostgreSQL full-text retrieval as a pragmatic design choice. Since the RAG corpus is small and curated, stage-filtered retrieval can return explainable guidance chunks without a separate vector store. Keeping the knowledge base in the same PostgreSQL persistence layer as requirements, artifacts, approvals, and audit logs also supports traceability, governance, and access-control integration. We do not claim that SQL-based retrieval is generally preferable to vector-based retrieval. Future versions of TimeArch can support hybrid retrieval by combining structured SQL filtering with vector-based semantic search.

\subsection{Implementation Technologies}

TimeArch is implemented as a web-based prototype with a modular architecture that supports interactive workflows and controlled execution.
The frontend is built with React and TypeScript and uses Mermaid for live diagram rendering.
The backend is implemented as TypeScript edge functions running on the Deno runtime, which handle execution requests, context preparation, and persistence.
Orchestration is implemented in LangChain.js with schema-bound tool calling for structured output.
LLM calls are handled through an integration layer that connects the orchestration logic with the selected language models.
TimeArch uses an LLM gateway rather than a single fixed model. The current configuration assigns lighter models to short checks and high-frequency utility calls, uses a default model for most structured stage outputs, and reserves stronger reasoning models for consolidation and validation tasks. The prototype also includes an LLM Models view where authenticated users can inspect the available models, their intended use, and the stages or agents that use them, while administrators can register additional OpenAI-compatible or local model endpoints.
A PostgreSQL database with row-level security stores requirements, drivers, architecture artifacts, agent runs, stage approvals, audit logs, and the RAG knowledge base. 
Architectural inputs and generated artifacts are treated as project-scoped data, while LLM calls are routed through the server-side gateway so that model-provider access can be configured centrally.
Document generation produces DOCX and PDF deliverables from the accumulated artifacts.

\section{Feasibility Evaluation}
\label{Evaluation}

We conducted a case-based feasibility evaluation to examine whether TimeArch can support the proposed architecture lifecycle from heterogeneous requirements to architecture-related artifacts.
The evaluation was carried out by members of the project team using representative software systems from different domains, including Intelligent Transport Systems (ITS), IoT monitoring systems, and mobile applications such as MoodFlow (see \cite{RepPack}).
These systems were selected to cover diverse architectural contexts in terms of scale, complexity, input completeness, and domain-specific constraints.
The evaluation focused on five aspects: (i) clarity and usability of generated artifacts, (ii) consistency across lifecycle stages, (iii) traceability between requirements, architectural drivers, and design decisions, (iv) coordination among agents, and (v) relevance of validation results.
The purpose was to assess the technical and workflow feasibility of TimeArch as an industrially motivated prototype. Specifically, we examined whether TimeArch could execute the proposed lifecycle, transform heterogeneous inputs into architecture-related artifacts, maintain traceability across stages, and produce outputs that can be reviewed by architects. The project team conducted the evaluation to support controlled and repeatable assessment across the selected cases.

The selected systems were evaluated using input materials with varying levels of detail.
Some cases included detailed requirements, constraints, and assumptions provided through documents such as SRS and BRD files, while others relied on high-level or incomplete inputs collected through audio discussions, structured forms, or natural language descriptions.
Depending on the scenario, TimeArch generated Software Requirements Specifications (SRS), Software Architecture Documents (SAD), Architecture Assessment Reports (AAR), and Full Architecture Packages (FAP).
The generated documents and supporting evaluation materials are provided in the publicly available supplementary replication package~\cite{RepPack}.
This package includes the complete generated documentation for the evaluated case systems, including SRS, SAD, AAR, and full architecture package outputs. Although TimeArch is designed to support a four-phase lifecycle, the evaluation placed more emphasis on requirement definition and architecture design because these phases produce the main inputs and candidate architecture artifacts. Validation, assurance, delivery, and evolution were also exercised through assessment reports, documentation packages, implementation-oriented checks, and drift/evolution-oriented functions, but their use in long-term industrial delivery and maintenance settings remains part of future evaluation.
The main observations are summarized below.

\begin{itemize}
\item \textbf{Requirements.}
TimeArch was able to extract structured requirement elements, identify ambiguities and inconsistencies, and derive architectural drivers from heterogeneous inputs. Across the evaluated cases, requirement extraction and categorization were observed to be relatively consistent. However, audio-based intake occasionally mis-segmented long recordings, and very large requirement sets increased critic latency.
These observations indicate the need for improved audio chunking, scalable context management, and more efficient handling of large requirement sets.
\item \textbf{Architecture Design.} 
TimeArch generated candidate architectural solutions by recommending architectural styles, performing tradeoff analysis, and producing system decomposition, data models, and API specifications. 
The generated artifacts generally reflected the provided requirements and captured design rationale and considered alternatives. 
The logical and component views were usually more detailed than process and deployment views. 
We also observed that some Mermaid diagrams failed to render because of invalid syntax in generated node identifiers, and that some style recommendations needed stronger justification. 
These findings suggest that the design stage is feasible, but that diagram validation, view-specific prompting, and stronger rationale generation are needed to improve artifact quality.

\item \textbf{Validation and Assurance.}
TimeArch evaluated generated architectures through quality attribute assessment inspired by ATAM, risk analysis, traceability checking, and consistency checks.
Repeated executions on the same input sometimes produced variations in quality scores and risk severity, which indicates a repeatability limitation relevant to practitioner trust. TimeArch partly mitigates uncontrolled variation through schema-bound outputs, traceability links, audit logs, evaluator checks, and human approval gates, but more stable scoring calibration is still needed.
The validation mechanism combines structural checks, traceability checks, consistency checks, standards-informed requirement checks, evaluator-agent review, and human review. However, some checks are still more effective at identifying artifact presence, structural completeness, and traceability gaps than at fully assessing deeper architectural soundness. In some cases, the explanation behind individual scores was shallow. These observations indicate that the validation stage requires stronger evidence-grounded explanations, more stable scoring configurations, and deeper semantic checks.

\item \textbf{Delivery and Evolution.} 
TimeArch produced consolidated architecture documentation, including SRS, SAD, Architecture Assessment Report, and Full Architecture Package outputs. The system also supported implementation-oriented checks, alignment between design artifacts and generated outputs, and evolution-oriented functions such as drift detection, gap analysis, and evolution planning. 
At the same time, larger document generation was occasionally affected by Edge Function timeouts, exported PDF and DOCX diagrams sometimes lost visual fidelity, and drift detection in brownfield projects was sensitive to incomplete legacy artifacts.

\end{itemize}

Overall, the evaluation indicates that TimeArch can operationalize the proposed lifecycle and generate architecture-related artifacts across different input conditions.
The results also show that human review remains necessary, especially for assessing the adequacy of design rationale, validating quality attribute tradeoffs, and confirming architecture decisions.
The findings should be interpreted as feasibility evidence, not as a conclusive assessment of industrial effectiveness. We did not conduct a controlled baseline comparison against a skilled architect using a general-purpose LLM or a structured prompt template. However, TimeArch differs from such simpler alternatives through its explicit lifecycle structure, stage-level governance, traceability across artifacts, retrieval-based grounding, evaluator/challenger agents, and human approval gates. Future studies with external architects, predefined rating criteria, baseline GenAI workflows, and practitioner-oriented frameworks such as the Technology Acceptance Model or Task-Technology Fit are needed to assess artifact quality, usefulness, repeatability, task fit, and adoption in industrial settings.

\section{Implications and Lessons Learned}
\label{Implications}

\textbf{Toward Agentic Software Architecture Design}:
Software architecture is well suited to agentic AI because it involves multiple connected reasoning tasks rather than a single generation step.
This view is consistent with prior work in multi-agent software engineering, agent design patterns, and agent architectures~\cite{He2025LLMMultiAgentSE}.
TimeArch illustrates how software architecture design can be supported as a staged and coordinated process with explicit review and control.
For practitioners, this suggests that agentic workflows may help structure architectural work while preserving human oversight.
At the same time, such support should strengthen, not replace, the role of architects.

\textbf{Explainability and Traceability of Outputs}:
As AI-generated architectural outputs become more detailed, their practical value depends on how well they can be explained, traced, and reviewed.
In software architecture, requirements, design decisions, tradeoffs, risks, and validation results must remain connected across stages rather than being treated as isolated artifacts~\cite{bass2012}.
Explainability therefore depends not only on readable outputs but also on showing how architectural reasoning progresses from requirements to decisions~\cite{Ozkaya2023ArchitectureKnowledgeGenAI}.
The feasibility evaluation suggests that TimeArch supports traceability across requirements, architectural drivers, design outputs, and validation artifacts.
However, traceability alone is not sufficient, since generated views, diagrams, and recommendations must also be understandable in practice.
This is consistent with prior studies showing that AI-supported architecting and modeling outputs still require careful review and interpretation~\cite{Camara2023ModelingTasksChatGPTUML}.

\textbf{Current State of TimeArch}:
Based on the feasibility evaluation, TimeArch provides initial support for software architecture design as a lifecycle-oriented and agent-based process.
The strongest observations concern requirement structuring, architectural driver identification, traceability, validation support, and consolidated architecture documentation.
These observations suggest that the approach can be useful as an initial environment for architecture drafting and review.
At the same time, some generated artifacts require improvements in clarity, readability, repeatability, and presentation.
The current prototype supports the main architecting stages, while still requiring refinement and more evaluation before stronger claims can be made about effectiveness or generalizability.

\textbf{Lessons Learned}:
The feasibility evaluation suggests four main lessons.
First, lifecycle-oriented architecting benefits from governed orchestration, since explicit stage structure, validation checks, and approval gates help maintain consistency and traceability.
Second, architecture generation benefits from evaluator-style review and human oversight, because they help identify risks and tradeoff concerns that may otherwise remain implicit.
Third, reliable AI-supported architecting depends on grounding and structural control; domain-specific grounding and structured schemas help maintain pipeline stability, but output stability and scoring consistency still require further refinement.
Fourth, traceability is useful only when generated artifacts are also understandable and reviewable, which highlights the need for better explanation, visualization, and artifact usability.

\section{Threats to Validity}
\label{ThreatstoValidity}

\textbf{Internal validity}: Project team members evaluated the system, which may introduce bias in how the generated artifacts were reviewed and interpreted. 
Although the evaluation considered criteria such as lifecycle coverage, artifact quality, consistency, traceability, validation support, and usability, some observations remain dependent on expert judgment. 
In addition, TimeArch relies on LLM-based components, which may introduce variability across runs. 
This variability may affect generated artifacts, validation scores, risk severity levels, and explanatory depth. 
To reduce this effect, the system uses staged execution, structured prompts, schema-bound outputs, and controlled data flow between phases. 
However, inconsistent or unexpected outputs cannot be fully ruled out.

\textbf{Construct validity}: TimeArch is intended to support software architecture design through generated requirements, architectural drivers, design views, validation outputs, and documentation. 
However, the extent to which these artifacts fully represent sound architectural reasoning remains a validity concern. 
The system uses RAG, domain-specific grounding, and structured output schemas to improve relevance and structural consistency, but these mechanisms do not guarantee that the generated artifacts fully capture the intended architectural concepts or decisions. 
In particular, some outputs may appear structurally complete while still lacking depth, precision, or contextual appropriateness.

\textbf{External validity}: The evaluation was conducted on a limited number of case systems from selected domains, including Intelligent Transport Systems, IoT monitoring systems, and mobile applications. 
Although these cases cover different architectural characteristics, they do not represent the full range of system types, scales, or industrial constraints. 
In addition, the evaluation was conducted in a controlled setting rather than in live industrial projects. 
Variations in requirement quality, organizational context, domain knowledge, available architecture expertise, and development practices may affect the applicability of the approach. Security and confidentiality are also relevant concerns for industrial use. Architecture inputs may include sensitive business requirements, legacy documentation, system constraints, and design rationale. Although TimeArch uses project-scoped storage, row-level security, server-side model routing, and separation between the RAG knowledge base and project-specific artifacts, these mechanisms do not remove all privacy risks. In particular, LLM inference may expose prompt content to model providers, and generated artifacts may reproduce sensitive input details. Future deployments should define organization-specific policies for provider selection, data retention, access control, and privacy review.

\textbf{Conclusion validity}: The conclusions are based on qualitative observations rather than statistical analysis or controlled comparison with other approaches. 
The evaluation thus supports claims about feasibility and observed strengths and limitations, but does not provide comparative evidence of effectiveness relative to other methods. 
To improve transparency, the generated artifacts and supporting evaluation materials are provided in the supplementary package~\cite{RepPack}.
\section{Related Work}
\label{RelatedWork}

\subsection{GenAI, LLMs, and the Architecture Gap}

GenAI has become important in SE, particularly for code generation, documentation, test assistance, and developer support~\cite{Ebert2023GenerativeAI,ernst2022ai}.
While this literature establishes the practical relevance of AI-assisted development, it remains largely implementation-oriented.
Ozkaya argues that architecture knowledge must remain central in AI-assisted development and asks how architectural concerns, constraints, and design rationale can guide GenAI systems~\cite{Ozkaya2023ArchitectureKnowledgeGenAI}.
This highlights a key limitation of current GenAI work in SE: it demonstrates utility for development tasks, but does not provide a disciplined approach for software architecture design, validation, and governance.

Work on LLM-assisted architecting and modeling provides evidence that LLMs can support design-related activities.
Ahmad et al. show that ChatGPT can assist architectural analysis, synthesis, and evaluation in a human--bot collaborative setting~\cite{Ahmad2023HumanBotArchitecting}, while C'{a}mara et al. show that generative AI can support UML-related modeling tasks, although output quality remains uneven and requires careful review~\cite{Camara2023ModelingTasksChatGPTUML}.
Hence, LLMs can contribute to early design exploration, modeling, and architecture-related reasoning.
However, the support is largely assistive and artifact-specific, and does not yet structure software architecture as a governed lifecycle with traceability from requirements to architectural drivers, validation checkpoints, and controlled progression toward implementation.

\subsection{Agentic and Multi-Agent Approaches}

Recent work has begun to provide architectural foundations for foundation-model-based agents.
Lu et al. propose a reference architecture for such agents and incorporate responsible-AI concerns such as accountability and security into the architectural perspective~\cite{Lu2024ResponsibleGenAIArchitecture}.
Liu et al. present an agent design pattern catalogue that distills reusable architectural patterns and associated trade-offs~\cite{Liu2025AgentPatternCatalogue}.
These contributions move beyond ad hoc LLM orchestration toward more systematic architectural thinking for agentic systems.

At the software-engineering level, He et al. show that LLM-based multi-agent systems improve task decomposition, coordination, and scalability across SE activities~\cite{He2025LLMMultiAgentSE}.
A smaller but highly relevant cluster focuses directly on software architecture automation.
Becattini et al. propose SALLMA, a software architecture for LLM-based multi-agent systems~\cite{Becattini2025SALLMA}.
Zhang et al. introduce a knowledge-based multi-agent framework for automated software architecture design~\cite{Zhang2025KnowledgeBasedMAAD}, and Li et al. extend this direction with MAAD, where specialized agents interpret requirements and generate architecture blueprints enriched with evaluation outputs~\cite{Li2025MAAD}.
These works provide the closest prior art to TimeArch, but remain limited in lifecycle coverage, focusing mainly on architecture generation, blueprinting, or partial evaluation rather than a governed end-to-end architecting process.

\subsection{Synthesis}

Existing work shows that GenAI supports implementation-oriented SE tasks, LLM-assisted design exploration, and multi-agent coordination~\cite{Ebert2023GenerativeAI,ernst2022ai,Ahmad2023HumanBotArchitecting,Camara2023ModelingTasksChatGPTUML,He2025LLMMultiAgentSE}.
Agent architecture research contributes reference architectures, taxonomies, and patterns for foundation-model-based agents, while architecture automation work shows that agents can support parts of architecture generation~\cite{Lu2024ResponsibleGenAIArchitecture,Liu2025AgentPatternCatalogue,Zhang2025KnowledgeBasedMAAD,Li2025MAAD,Becattini2025SALLMA}.
However, these streams still provide limited support for a governed, lifecycle-oriented architecting process that connects requirements, architectural drivers, validation checkpoints, traceability, and architecture-aligned delivery.
TimeArch addresses this by combining specialized agents, retrieval-grounded architectural knowledge, validation checkpoints, and traceable decision support from requirements to validated architectural artifacts and architecture-aligned delivery.

\section{Conclusion}
\label{Conclusion}

This paper presented \textbf{TimeArch}, a multi-agent approach and working prototype for lifecycle-oriented software architecture design. 
TimeArch introduces a governed architectural layer between requirements and implementation by combining generator and evaluator agents, knowledge grounding, traceability, and human approval checkpoints.

The case-based feasibility evaluation suggests that TimeArch can support structured architecting and produce initial architectural artifacts for human review. 
The results also show that human judgement remains necessary for assessing design rationale, quality-attribute tradeoffs, and architectural decisions.

Future work will focus on improving the clarity, consistency, repeatability, and presentation of generated outputs. 
We also plan to conduct comprehensive industrial case studies with companies to evaluate the adoption and use of TimeArch in both new software projects and existing systems, including comparisons with baseline GenAI workflows.

\begin{credits}
\subsubsection{\ackname} This work has been supported by Business Finland (project ANSE (1822/31/2025)).

\subsubsection{Acknowledgement of AI Tools.}

\noindent AI tools were used during the preparation of this work. Lovable was used to assist code generation in the implementation of TimeArch. Generated outputs were reviewed, revised, and validated before inclusion in the system. ChatGPT was used for language refinement and it did not contribute to the scientific content. Responsibility for the manuscript remains with the authors.
\end{credits}
\bibliographystyle{splncs04}
\bibliography{refs}

\end{document}